\documentstyle[12pt]{article}
\input epsf.sty
 1

\catcode`\@=11 % This allows us to modify PLAIN macros.
\makeatletter
\def\@seccntformat#1{\csname the#1\endcsname.\hskip 1em}
\makeatother
\begin{document}

%%%%%%%%%%%%%%%%%%%%%%%%%%%%%%%%%%%%%%%%%%%%%%%%%%%%%%%
%------------------------------------------------------
% Title page
%------------------------------------------------------
\thispagestyle{empty}
\begin{flushright}

{\footnotesize\renewcommand{\baselinestretch}{.75}
           SLAC-PUB-8408\\
           June 2000\\
}
\end{flushright}

\vspace {0.5cm}

\begin{center}
{\large \bf  FIRST DIRECT MEASUREMENT OF THE PARITY-VIOLATING 
COUPLING OF THE Z$^{0}$ TO THE $s$-QUARK$^*$}

\vspace {1.5cm}

 {\bf The SLD Collaboration$^{**}$}

\vspace {0.5cm}

Stanford Linear Accelerator Center \\
Stanford University, Stanford, CA~94309 \\

\vspace{2.5cm}
{\bf Abstract}
\end{center}

\renewcommand{\baselinestretch}{1.2}

We have made the first direct measurement of the parity-violating coupling of
the $Z^0$ boson to the strange quark, $A_s$, using $\sim$550,000
$e^+e^-\!\rightarrow \!Z^0\!\!\rightarrow$hadrons events produced with a
polarized electron beam and recorded by the SLD experiment.
$Z^0\!\!\rightarrow \!s\bar{s}$ events were tagged by the
absence of $B$ or $D$ hadrons and
the presence in each hemisphere of a high-momentum $K^\pm$ or $K^0_s$.
From the polar angle distributions of the strangeness-signed thrust axis, we
obtained 
$A_s \!=\!0.895\!\pm\!0.066 (stat.)\!\pm\!0.062 (syst.)$.
The analyzing power and
$u\bar{u}\!+\!d\bar{d}$ background were constrained using the data.

\vspace{3.3cm}
\begin{center}
{\it Submitted to Physical Review Letters.}
\end{center}

\vfil

\noindent
$^*$Work supported in part by Department of Energy contract DE-AC03-76SF00515.
\eject

%%\section{Introduction}

The extent of parity violation in the electroweak coupling of the $Z^0$ boson
to an elementary fermion $f$ can be specified by the parameter
$A_f\!=\!2v_f a_f/(v_f^2\!+\!a_f^2)$, where $v_f$ ($a_f$) is the vector
(axial-vector) $Zf\bar{f}$ coupling.
In the Standard Model (SM), universal couplings are expected for the 
leptons ($A_e\!=\!A_\mu\!=\!A_\tau$),
the down-type quarks ($A_d\!=\!A_s\!=\!A_b$) and
the up-type quarks ($A_u\!=\!A_c\!=\!A_t$).
Precise measurements of the $A_f$ allow stringent tests of the SM and
sensitivity through radiative corrections to e.g.:
the top quark and Higgs boson masses ($A_{e,\mu,\tau}$);
new physics that affects primarily the right-handed couplings ($A_{d,s,b}$);
and
new physics that couples more strongly to heavier quarks (deviations from
universality).

All except $A_t$ can be measured in $e^+e^-$ annihilations at the $Z^0$
resonance via forward-backward production asymmetries in $\theta_f$, the polar
angle of the outgoing $f$ with respect to the incoming $e^-$ beam.
At the SLC, the $e^-$ beam has longitudinal polarization $P_e$, the $e^+$
beam is unpolarized, and the Born-level differential cross-section for the
process $e^+e^-\!\rightarrow\!Z^0\!\!\rightarrow\!f\bar{f}$ is:
\begin{equation}
d\sigma_f / dx \propto (1\!-\!A_e P_e)(1\!+\!x^2) + 2A_f(A_e\!-\!P_e)x ,
\end{equation}
where the last term is antisymmetric in $x\!=\!\cos\theta_f$.
Using both left- ($P_e\!<\!0$) and right-polarized
($P_e\!>\!0$) beams of magnitude $|P_e|$, one can measure both the initial-
($A_e$) and final-state ($A_f$) couplings directly \cite{epol,sldab};
for $P_e\!=\!0$ one can measure only their product,
or $A_{FB}^f \equiv 3 A_e A_f/4$.

The most precisely measured coupling is $A_e$, with a relative error of
1.3\%~\cite{epol,ewlepslc}, and lepton universality is verified at the 8\%
level~\cite{ewlepslc}.
In the quark sector, several measurements of $A_b$ and $A_c$ that use properties
of the leading $B$ and $D$ hadrons can be combined to yield precisions of 2.0\%
and 4.4\%, respectively~\cite{ewlepslc}.
However there are few measurements of $A_u$, $A_d$ or $A_s$ \cite{opal,delphi}
because the leading particles in $u$, $d$ and $s$ jets are more difficult to
identify experimentally; they have relatively low energy, are not unique to
events of a particular flavor, and nonleading particles of the same species are
produced in hadronic jets of all flavors.
Furthermore, these aspects of jet fragmentation are not well measured, and
previous indirect measurements either relied on imprecise constraints from
their data (OPAL: $A^u_{FB}\!=\!0.040\!\pm\!0.073$;
$A^{ds}_{FB}\!=\!0.068\!\pm\!0.037$~\cite{opal}) or
are model-dependent (DELPHI: $A^s_{FB}\!=\!0.101\!\pm\!0.012$~\cite{delphi}).

In this Letter we present the first direct measurement of $A_s$.
We used high-momentum $K^\pm$ and $K^0_s$ to tag $Z^0\!\!\rightarrow\!s\bar{s}$
events, and the $K^\pm$ charge to separate $s$ jets from $\bar{s}$ jets.
The heavy flavor ($c\bar{c}\!+\!b\bar{b}$) event background was suppressed by
identifying $B$ and $D$ decay vertices.
The $u\bar{u}\!+\!d\bar{d}$ background was suppressed and the $s$-$\bar{s}$
separation enhanced by requiring an $s$/$\bar{s}$-tag in each event hemisphere,
reducing any model dependence.
The remaining $u\bar{u}\!+\!d\bar{d}$ background and the $s$-$\bar{s}$
separation were constrained using related observables in the data.

%%\section{Apparatus and Hadronic Event Selection}

We used the sample of approximately 550,000 hadronic $Z^0$ decays recorded by 
the SLD~\cite{sld} experiment at the SLAC Linear Collider, with
$\left< |P_e| \right>\!=\!0.735\pm 0.005$~\cite{epol}, from 1993--1998.
Charged tracks were measured in the Central Drift Chamber (CDC)~\cite{cdc} and
the original (upgraded) Vertex Detector (VXD)~\cite{vxd} in 26.5\% (73.5\%) of
the data;
the resolution on the impact parameter $d$ in the plane
perpendicular to the beam direction, including the uncertainty on the
interaction point, was 
$\sigma_d =$11$\oplus$70/$(p \sin^{3/2}\theta)$ $\mu$m
(8$\oplus$29/$(p \sin^{3/2}\theta)$ $\mu$m),
where $p$ is the track momentum in GeV/c and $\theta$ its polar angle with
respect to the beamline.
Tracks were identified as $\pi^\pm$, $K^\pm$ or p/$\bar{\rm p}$
in the Cherenkov Ring Imaging Detector (CRID)~\cite{crid}, which
allowed the identification with high efficiency and purity
of $\pi^\pm$ with $0.3\!<\!p\!<\!35$ GeV/c, $K^\pm$ with
$0.75\!<\!p\!<\!6$ GeV/c or $9\!<\!p\!<\!35$ GeV/c, and p/$\bar{\rm p}$ with
$0.75\!<\!p\!<\!6$ GeV/c or $10\!<\!p\!<\!46$ GeV/c~\cite{pprod}.
The event thrust axis~\cite{thrust} was calculated using energy clusters
measured in the Liquid Argon Calorimeter~\cite{lac}.

After selecting hadronic $Z^0$ decays~\cite{homer}, we removed $c\bar{c}$ and
$b\bar{b}$ events by requiring no more than one well-measured~\cite{homer} track
with $d/\sigma_d\!>\!2.5$ in the event.
The efficiency for selecting light-flavor events with
$|\cos\theta_{thrust}|\!<$0.71
and the VXD, CDC and CRID operational was estimated to be over 95\%;
the selected sample comprised 205,708 events, with an estimated
contribution of 14.2\% (3.4\%) from $c\bar{c}$ ($b\bar{b}$) events.
Such performance parameters were estimated from a detailed 
Monte Carlo (MC) simulation \cite{homer,sldrb} of the SLD based on the
JETSET 7.4~\cite{jetset} event generator, tuned to reproduce many
measured properties of hadronic $Z^0$ decays, including the momentum-dependent
production of $K^\pm$, $K^0$, $K^*$ and $\phi$ mesons.

%%\section{Selection of $s\bar{s}$ Events}

Each selected event was divided into two hemispheres by the plane perpendicular
to the thrust axis, and in each hemisphere we searched for high-momentum strange
particles $K^\pm$, $K^0_s$ and $\Lambda^0/\bar{\Lambda}^0$.
Candidate $K^\pm$ tracks were required to have $p\!>\!9$ GeV/c, $d\!<\!1$ mm, 
to extrapolate through an active region of the CRID gaseous radiator system, and
to have a log-likelihood~\cite{pprod} for the $K^\pm$ hypothesis ${\cal L}_K$
that exceeded both ${\cal L}_\pi$ and ${\cal L}_{\rm p}$ by at least 3 units.
For $p\!>\!9$ GeV/c, the estimated $K^\pm$ selection efficiency (purity) was
48\% (91.5\%). 

Candidate $K_s^0\!\rightarrow\! \pi^+\pi^-$ and
$\Lambda^0$/$\bar\Lambda^0\!\rightarrow$p$\pi^-$/$\bar{\rm p}\pi^+$ decays were
reconstructed as described in \cite{pprod,hermann} from tracks not identified
as $K^\pm$.
We required $p\!>\!5$ GeV/c and a reconstructed invariant
mass $m_{\pi\pi}$ or $m_{{\rm p}\pi}$ within two standard deviations of the
$K_s^0$ or $\Lambda^0$ mass.
If CRID information was available for the p/$\bar{\rm p}$ track in a
$\Lambda^0$/$\bar\Lambda^0$ candidate, we required
${\cal L}_{\rm p} > {\cal L}_{\pi}$;
otherwise we required that the $\Lambda^0$/$\bar\Lambda^0$ not be a
$K^0_s$ candidate and that the flight distance exceed 10 times its uncertainty.
The estimated
%$\Lambda^0$/$\bar\Lambda^0\!\rightarrow\! {\rm p}\pi^-/\bar{\rm p}\pi^+$
$\Lambda^0$/$\bar\Lambda^0$ reconstruction
efficiency (purity) was 12\% (90.7\%).
These $\Lambda^0$/$\bar\Lambda^0$ were removed from the $K^0_s$
sample, for an estimated $K^0_s$ efficiency (purity)
of 24\% (90.7\%).

We considered only the selected strange particle with the highest momentum in
each hemisphere (5.5\% of those tagged contained more than one), and tagged the
event as $s\bar{s}$ if one hemisphere contained a $K^\pm$ and the other
contained either an oppositely charged $K^\pm$ or a $K^0_s$.
The $\Lambda^0$/$\bar{\Lambda}^0$ tags provided a useful veto in multiply
tagged hemispheres and important checks of the simulation;
however their inclusion did not improve the total error on $A_s$.
The thrust axis, signed so as to point into the hemisphere containing
(opposite) the $K^-$ ($K^+$), was used as an estimate of the initial $s$-quark
direction.
Table~\ref{dtags} shows the number of events tagged in each mode,
along with the predictions of the simulation, which are consistent.
Also shown are the simulated $s\bar{s}$ event purities and analyzing powers
$a_s\!\equiv\! (N_r\!-\!N_w)/(N_r\!+\!N_w)$, where $N_r$ ($N_w$) is the number
of events in which the signed thrust axis pointed into the true $s$ ($\bar{s}$)
hemisphere.

\begin{table}
\begin{center}
\begin{tabular}{|c|cc|cc|}
\hline
     & \# Events & MC         & $s \bar s$ & Analyzing \\[-0.1cm]
Mode & in Data   & Prediction & Purity     & Power \\
\hline
$K^+ K^-$                               & 1290 & 1312 & 0.73 & 0.95  \\
$K^\pm K_s^0$                           & 1580 & 1617 & 0.60 & 0.70  \\
\hline
$K^+ \Lambda^0, K^- \bar \Lambda^0$     &  219 &  213 & 0.66 & 0.89  \\
$\Lambda^0 \bar \Lambda^0$              &   17 &   14 & 0.57 & 0.70  \\
$\Lambda^0 K_s^0, \bar \Lambda^0 K_s^0$ &  193 &  194 & 0.50 & 0.32  \\
\hline
\end{tabular}
\caption{\label{dtags} \baselineskip=12pt
Summary of the selected event sample for the two tagging modes and the
three cross-check modes.
}
\end{center}
\end{table}

%%\section{Extraction of $A_s$}

\begin{figure}[t]
 \centering
 \epsfxsize16.1cm
 \epsffile{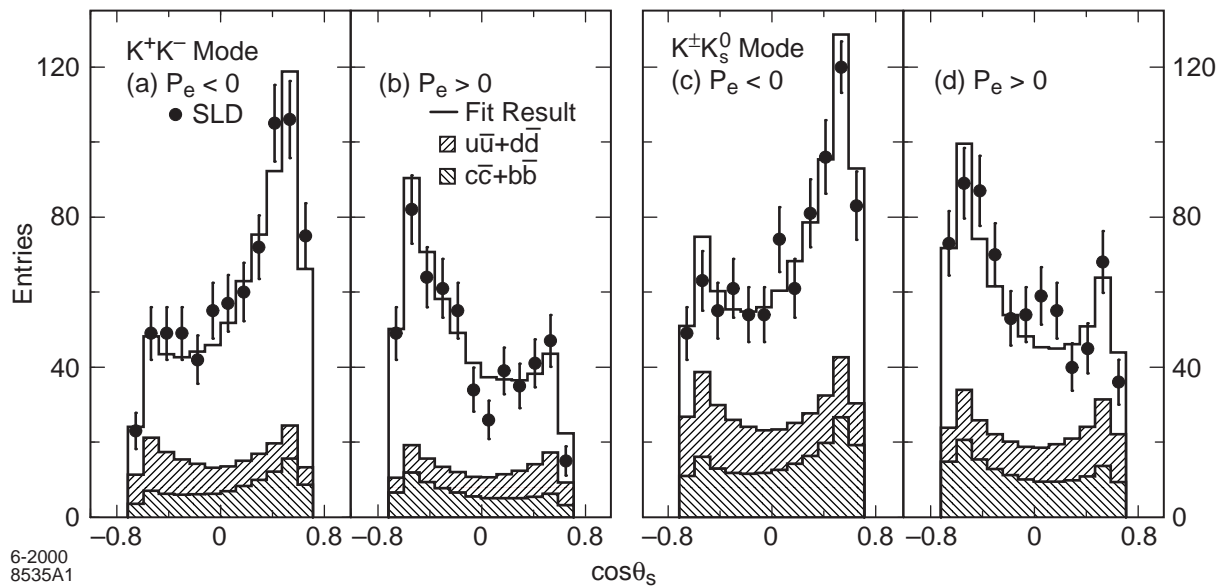}
 \caption{Measured $s$-quark polar angle distributions (dots) for selected
  events in the a,b) $K^+K^-$ and c,d) $K^\pm K^0_s$ modes, produced with
  a,c) left- and b,d) right-polarized electron beams. 
  The histograms represent the result of a simultaneous fit to the four
  distributions, and the upper (lower) hatched areas indicate the
  estimated $u\bar{u}\!+\!d\bar{d}$ ($c\bar{c}\!+\!b\bar{b}$) backgrounds.
 }
 \label{asymm}
 \end{figure}

Figure~\ref{asymm} shows the distributions of the measured $s$-quark polar
angle $\theta_s$ for the $K^+K^-$ and $K^\pm K^0_s$ modes.
In each case, production asymmetries of opposite sign and different magnitude
for left- and right-polarized $e^-$ beams are visible.
The content of the largest $|\cos\theta_s|$ bins is reduced by the detector
acceptance.
The estimated backgrounds (discussed below) are indicated:  those from
$c\bar{c}\!+\!b\bar{b}$ events exhibit asymmetries of the same
sign and similar magnitude to those of the signal, so the measured $A_s$ is
largely insensitive to them;
those from $u\bar{u}\!+\!d\bar{d}$ events exhibit asymmetries of
opposite sign, and $A_s$ is more sensitive to the associated uncertainties.

A simultaneous maximum likelihood fit to these four distributions was
performed using the function:
\begin{equation}
    L = \prod_{k = 1}^{N_{data}} \sum_{q=udscb}^{} N_q 
          \{ (1\!-\!A_e P_e)(1\!+\!x^2_k) +     
       2 (A_e\!-\!P_e) (1\!+\!\delta) a_q A_q x_k\}.  \label{likfcn}
\end{equation} 
Here, the number of tagged $q\bar{q}$ events $N_q\!=\!N_{events}R_q\epsilon_q$,
$R_q\!=\!\Gamma(Z^0\!\!\rightarrow \!q\bar{q})/\Gamma(Z^0\!\!\rightarrow$hadrons$)$,
$\epsilon_q$ is the tagging efficiency,
$a_q$ is the analyzing power for tagging the $q$ direction, and the
correction for hard gluon radiation $\delta\!=\!-0.013$ was
derived \cite{shinya} as in~\cite{sldab}.
The values of the $\epsilon_q$ and $a_q$ depend on the tagging mode.
World average values~\cite{ewlepslc} of $A_e$, $A_c$, $A_b$,
$R_c$ and $R_b$ were used, along with SM values of $A_u$, $A_d$, $R_u$,
$R_d$ and $R_s$.
Simulated values of $\epsilon_c$, $\epsilon_b$, $a_c$ and $a_b$ were used,
as they depend primarily on measured quantities with well defined uncertainties.

For the light flavors, the relevant parameter values were derived where
possible from the data.
The number of events $N_u\!+\!N_d\!+\!N_s$ was determined by subtracting
the simulated $N_c$ and $N_b$ from the total observed.
The values of $a_s$ and the ratio $(N_u\!+\!N_d)/N_s$ were constrained
(see below) using the data;
since the simulation was consistent with the data, the simulated values of
$a_s$ were used and the simulated $\epsilon_u$, $\epsilon_d$ and $\epsilon_s$
were scaled by a common factor to give the measured $N_u\!+\!N_d\!+\!N_s$.
The average $a_{ud}\equiv (N_u a_u\!+\!N_d a_d)/(N_u\!+\!N_d)$ can also be
constrained from the data; however our constraint is less precise than the
range $-a_s\!<\!a_u,a_d\!<\!0$, obtained by noting that a $u$ ($d$) jet can
produce a leading $K^+$ ($K^{*0}\!\rightarrow\!K^+\pi^-$), giving
$a_u$($a_d$)$<$0, but with an associated $K^-$ or $\bar{K}^0$, and a $K^-$
can be selected with reduced probability, giving $|a_u|$($|a_d|$)$<\!|a_s|$.
We scaled the simulated $a_u$ and $a_d$ by a common factor such that
$a_{ud}\!=\!-a_s/2$ for each mode.
The fit yielded $A_s\!=\!0.895\!\pm\!0.066$ (stat.).
Histograms corresponding to this value are shown in fig.~\ref{asymm} and are 
consistent with the data; the binned $\chi^2$ is 42 for 48 bins.

%%\section{Systematic Uncertainties and Checks}

We considered several sources of systematic uncertainty, summarized
in table~\ref{systerr}.
The values of $R_c$, $R_b$, $A_c$ and $A_b$ were varied by the uncertainties
on their world averages~\cite{ewlepslc}.
A large number of quantities in the simulation of heavy flavor events and
detector performance were varied as in \cite{sldrb} with
negligible effect on the measured value of $A_s$.
The yield and analyzing power of true $K^\pm$ from $D$ ($B$) decays
have been derived from SLD data in the context of a measurement \cite{twright}
of $A_c$ ($A_b$), and our simulation reproduces them within the measurement
errors.
We applied corresponding relative variations of $\pm$9\% to $\epsilon_c$,
$\pm$3.3\% to $\epsilon_b$, $\pm$5\% (15\%) to $a_c$ and $\pm$3.6\% (4.4\%)
to $a_b$ for the $K^+K^-$ ($K^\pm K^0_s$) mode.
The sum in quadrature of the uncertainties due to heavy flavor background
is listed in Table~\ref{systerr};
the largest contribution is from $\delta a_c$.

The key to this measurement is the understanding of the light-flavor parameters,
for which there are few experimental constraints, and these gave rise to the
dominant systematic uncertainties~\cite{hermann}.
In order to minimize model dependence, we used our data to constrain the largest
uncertainties in these parameters within the context of our
simulation,
which reproduces existing measurements of relevant quantities such as leading
particle production and strange-antistrange correlations~\cite{hermann}.

To constrain the analyzing power $a_s$, we note that to mistag an $s$ jet as
an $\bar{s}$ jet we must either identify a true $K^+$ or misidentify a $\pi^+$
or p as a $K^+$.
A true high-momentum $K^+$ in an $s$ jet must be produced in association
with an antistrange particle, yielding a jet with three high-momentum particles
of nonzero strangeness.
In our data we found 61 hemispheres containing three selected $K^{\pm}$
and/or $K_s^0$; the MC prediction of 67 is consistent.
We quantified this as a constraint on $a_s$ by subtracting the simulated
($c\!+\!\bar{c}\!+\!b\!+\!\bar{b}$) contribution of 9.3,
scaling by the simulated ratio
$(s\!+\!\bar{s})/(u\!+\!\bar{u}\!+\!d\!+\!\bar{d}\!+\!s\!+\!\bar{s})=0.74$, and
comparing with the MC prediction for ($s\!+\!\bar{s})$.
Propagating the data and MC statistical errors yielded an 18.5\% relative
constraint on the wrong-sign fraction, $w_s=(1\!-\!a_s)/2$, in $s$/$\bar{s}$
hemispheres.
This constraint is not entirely model-independent
but any further uncertainties are small compared with 18.5\%.
Assuming equal production of charged and neutral kaons, this procedure
delivers a calibration of $w_s$ for both tagging modes,
which we varied simultaneously by $\pm$18.5\% relative.
To account for misidentified particles we varied the production of $>$9 GeV/c 
p and $\pi^+$ in $s$ jets by $\pm$100\%, and varied the
misidentification probability by its measured relative uncertainty of
$\pm$25\% \cite{pprod}.
The sum in quadrature of these three effects is shown in table \ref{systerr} and
is dominated by the 3-kaon calibration.

The relative $u\bar{u}\!+\!d\bar{d}$ background $B_{ud}=(N_u\!+\!N_d)/N_s$ was
constrained in a similar manner, by exploiting the fact that an even number
of particles with nonzero strangeness must be produced in a $u$ or $d$ jet.
The three quantities, the number $N_1\!=\!1262$ of hemispheres in the
data containing an identified $K^+K^-$ pair, $N_2\!=\!983$ hemispheres
containing a $K^\pm K^0_s$ pair, and $N_3\!=\!503$ events with an identified
$K^\pm$ of the same charge in both hemispheres, constrain $B_{ud}$ in
complementary ways: $N_1$ and $N_2$ are primarily sensitive to $K\bar{K}$
production in $u$/$d$ jets; $(N_1\!-\!N_2)$ to $\phi$ production in $s$ jets;
and $N_3$ to these and also the production and misidentification of $\pi^\pm$
and p/$\bar{\rm p}$.
Furthermore, all are sensitive to deviations from the assumed values of $R_u$,
$R_d$ and $R_s$.
The MC predictions of $N_1\!=\!1218$, $N_2\!=\!1002$ and $N_3\!=\!559$ are
consistent, and
relative constraints on $B_{ud}$ of 4.6\%, 5.1\% and 8.1\%, respectively,
were derived.
Since $N_3$ constrains the sum of all contributions,
we varied $B_{ud}$ by $\pm$8.1\%.

These quantities are also sensitive to $a_u$ and $a_d$,
however our limited event sample did not allow us to obtain a useful constraint.
We therefore took $-a_s\!<\!a_u,a_d\!<\!0$ as hard limits and
scaled $a_u$ and $a_d$ simultaneously such that 
$a_{ud}\!=\!-a_s/2\!\pm\!a_s/\sqrt{12}$.
This yielded the dominant systematic error on $A_s$ and is a quantity that must
be understood experimentally before a more precise measurement can be made.
Since the product $A_q a_q$ appears in Eqn. \ref{likfcn}, this is
equivalent to varying $A_u$ and $A_d$ down to half of their SM values and
up to well over unity; we considered no additional variation of $A_u$ or $A_d$.
The uncertainties listed in table \ref{systerr} were added in quadrature to
yield a total relative systematic error of $\pm$0.069.

\begin{table}
\begin{center}
\begin{tabular}{|ll|r@{$\pm$}l|c|}
\hline
         &  & \multicolumn{2}{c|}{Systematic}&      \\[-0.1cm]
Source   &  & \multicolumn{2}{c|}{variation} & $\delta A_s / A_s$ \\
\hline
\multicolumn{2}{|l|}{Heavy flavor background}                 &
\multicolumn{2}{|c|}{See text}                                & 0.014 \\
\hline 
\multicolumn{2}{|l|}{Correction for gluon radiation}
                                        &$-0.013$ & $0.006$   & 0.006 \\
\hline 
$\left< |P_e| \right>$  &               & $0.735$ & $0.005$   & 0.006 \\
\hline
MC statistics           &              &\multicolumn{2}{c|}{ }& 0.014 \\
\hline 
$a_s$                   & $K^+K^-$      & $0.949$ & $0.012$   & \\[-0.1cm]
                        & $K^\pm K^0_s$ & $0.701$ & $0.060$   & 0.032 \\
\hline
$(N_u+N_d)/N_s$         & $K^+K^-$      & $0.190$ & $0.015$   & \\[-0.1cm]
(incl. $(R_u+R_d)/R_s$) & $K^\pm K^0_s$ & $0.316$ & $0.026$   & 0.021 \\
\hline 
$a_u, a_d$              &         & $-a_s/2$ & $a_s/\sqrt{12}$ & 0.054 \\[-0.1cm]
$A_u, A_d$              &               &\multicolumn{2}{c|}{--}&     \\
\hline \hline
Total                   &               &\multicolumn{2}{c|}{ }& 0.069 \\
\hline
\end{tabular}
\caption{\label{systerr} Summary of the systematic uncertainties.
}
\end{center}
\end{table}

Several systematic checks were also performed.
Ad hoc corrections~\cite{hermann} to the simulation of the kaon momentum
distributions and
identification efficiencies, and the charged track reconstruction efficiency and
impact parameter resolution were removed and the analysis repeated;
changes in the measured value of $A_s$ were much smaller than the systematic
error.
We fitted each tagging mode separately, including those involving $\Lambda^0$
tags, with consistent results.
We repeated the analysis using all $K^\pm$, and all $\Lambda^0/\bar{\Lambda}^0$,
hemispheres with no tag required in the opposite hemisphere; results were
consistent.
This $K^\pm$ analysis is similar to that in \cite{delphi}; it has a relative
statistical precision of 0.03, but of 0.18 systematic.

%%\section{Summary and Conclusion}

In conclusion, we have made the first direct measurement of the
parity-violating coupling of the $Z^0$ boson to the strange quark,
\[
A_s \!=\! 0.895 \!\pm\! 0.066 (stat.) \pm 0.062 (syst.),
\]
using high-momentum identified $K^\pm$ and $K^0_s$ to tag
$Z^0\!\!\rightarrow\!s\bar{s}$ decays and determine the $s$-quark direction.
Our high $K^\pm$ identification efficiency allowed the use of a relatively
high-purity, double-tagged event sample, and the extraction from the data of
constraints on the analyzing power of the method and the
$u\bar{u}\!+\!d\bar{d}$ background, using events with same-charge
double tags and jets with two or three identified kaons.
This result is consistent with the Standard Model expectation, $A_s\!=\!0.935$,
and with less precise, previous measurements of $A_{FB}^s$ \cite{opal,delphi}.
It is also consistent with a recent world average $b$-quark asymmetry,
$A_b\!=\!0.881\!\pm\!0.018$~\cite{ewlepslc},
providing a 10\% test of down-type quark universality.

%%\section*{Acknowledgements}
We thank the personnel of the SLAC accelerator department and the
technical
staffs of our collaborating institutions for their outstanding efforts
on our behalf.
This work was supported by the U.S. Department of Energy, 
  the UK Particle Physics and Astronomy Research Council
  (Brunel, Oxford and RAL);
  the Istituto Nazionale di Fisica Nucleare of Italy
  (Bologna, Ferrara, Frascati, Pisa, Padova, Perugia);
  the Japan-US Cooperative Research Project on High Energy Physics
  (Nagoya, Tohoku);
  and the Korea Science and Engineering Foundation (Soongsil).

\section*{$^{**}$List of Authors}
%
% author list for inclusion in LaTeX documents
% using \author{} and \address{} commands
%
% Instion number definitions:
%
\begin{center}
\def\iAOMORI{$^{(1)}$}
\def\iBRI{$^{(2)}$}
\def\iBRUN{$^{(3)}$}
\def\iBU{$^{(4)}$}
\def\iCOLO{$^{(5)}$}
\def\iCSU{$^{(6)}$}
\def\iFERR{$^{(7)}$}
\def\iFRAS{$^{(8)}$}
\def\iJHU{$^{(9)}$}
\def\iLBL{$^{(10)}$}
\def\iMASS{$^{(11)}$}
\def\iMISSI{$^{(12)}$}
\def\iMIT{$^{(13)}$}
\def\iMOSCOW{$^{(14)}$}
\def\iNAGO{$^{(15)}$}
\def\iOREG{$^{(16)}$}
\def\iOXF{$^{(17)}$}
\def\iPERU{$^{(18)}$}
\def\iRAL{$^{(19)}$}
\def\iRUTG{$^{(20)}$}
\def\iSLAC{$^{(21)}$}
\def\iSOONG{$^{(22)}$}
\def\iTENN{$^{(23)}$}
\def\iTOHO{$^{(24)}$}
\def\iUCSB{$^{(25)}$}
\def\iUCSC{$^{(26)}$}
\def\iVAND{$^{(27)}$}
\def\iWASH{$^{(28)}$}
\def\iWISC{$^{(29)}$}
\def\iYALE{$^{(30)}$}

  \baselineskip=.75\baselineskip
\mbox{Koya Abe\unskip,\iTOHO}
\mbox{Kenji Abe\unskip,\iNAGO}
\mbox{T. Abe\unskip,\iSLAC}
\mbox{I. Adam\unskip,\iSLAC}
\mbox{H. Akimoto\unskip,\iSLAC}
\mbox{D. Aston\unskip,\iSLAC}
\mbox{K.G. Baird\unskip,\iMASS}
\mbox{C. Baltay\unskip,\iYALE}
\mbox{H.R. Band\unskip,\iWISC}
\mbox{T.L. Barklow\unskip,\iSLAC}
\mbox{J.M. Bauer\unskip,\iMISSI}
\mbox{G. Bellodi\unskip,\iOXF}
\mbox{R. Berger\unskip,\iSLAC}
\mbox{G. Blaylock\unskip,\iMASS}
\mbox{J.R. Bogart\unskip,\iSLAC}
\mbox{G.R. Bower\unskip,\iSLAC}
\mbox{J.E. Brau\unskip,\iOREG}
\mbox{M. Breidenbach\unskip,\iSLAC}
\mbox{W.M. Bugg\unskip,\iTENN}
\mbox{D. Burke\unskip,\iSLAC}
\mbox{T.H. Burnett\unskip,\iWASH}
\mbox{P.N. Burrows\unskip,\iOXF}
\mbox{A. Calcaterra\unskip,\iFRAS}
\mbox{R. Cassell\unskip,\iSLAC}
\mbox{A. Chou\unskip,\iSLAC}
\mbox{H.O. Cohn\unskip,\iTENN}
\mbox{J.A. Coller\unskip,\iBU}
\mbox{M.R. Convery\unskip,\iSLAC}
\mbox{V. Cook\unskip,\iWASH}
\mbox{R.F. Cowan\unskip,\iMIT}
\mbox{G. Crawford\unskip,\iSLAC}
\mbox{C.J.S. Damerell\unskip,\iRAL}
\mbox{M. Daoudi\unskip,\iSLAC}
\mbox{S. Dasu\unskip,\iWISC}
\mbox{N. de Groot\unskip,\iBRI}
\mbox{R. de Sangro\unskip,\iFRAS}
\mbox{D.N. Dong\unskip,\iMIT}
\mbox{M. Doser\unskip,\iSLAC}
\mbox{R. Dubois\unskip,}
\mbox{I. Erofeeva\unskip,\iMOSCOW}
\mbox{V. Eschenburg\unskip,\iMISSI}
\mbox{E. Etzion\unskip,\iWISC}
\mbox{S. Fahey\unskip,\iCOLO}
\mbox{D. Falciai\unskip,\iFRAS}
\mbox{J.P. Fernandez\unskip,\iUCSC}
\mbox{K. Flood\unskip,\iMASS}
\mbox{R. Frey\unskip,\iOREG}
\mbox{E.L. Hart\unskip,\iTENN}
\mbox{K. Hasuko\unskip,\iTOHO}
\mbox{S.S. Hertzbach\unskip,\iMASS}
\mbox{M.E. Huffer\unskip,\iSLAC}
\mbox{X. Huynh\unskip,\iSLAC}
\mbox{M. Iwasaki\unskip,\iOREG}
\mbox{D.J. Jackson\unskip,\iRAL}
\mbox{P. Jacques\unskip,\iRUTG}
\mbox{J.A. Jaros\unskip,\iSLAC}
\mbox{Z.Y. Jiang\unskip,\iSLAC}
\mbox{A.S. Johnson\unskip,\iSLAC}
\mbox{J.R. Johnson\unskip,\iWISC}
\mbox{R. Kajikawa\unskip,\iNAGO}
\mbox{M. Kalelkar\unskip,\iRUTG}
\mbox{H.J. Kang\unskip,\iRUTG}
\mbox{R.R. Kofler\unskip,\iMASS}
\mbox{R.S. Kroeger\unskip,\iMISSI}
\mbox{M. Langston\unskip,\iOREG}
\mbox{D.W.G. Leith\unskip,\iSLAC}
\mbox{V. Lia\unskip,\iMIT}
\mbox{C. Lin\unskip,\iMASS}
\mbox{G. Mancinelli\unskip,\iRUTG}
\mbox{S. Manly\unskip,\iYALE}
\mbox{G. Mantovani\unskip,\iPERU}
\mbox{T.W. Markiewicz\unskip,\iSLAC}
\mbox{T. Maruyama\unskip,\iSLAC}
\mbox{A.K. McKemey\unskip,\iBRUN}
\mbox{R. Messner\unskip,\iSLAC}
\mbox{K.C. Moffeit\unskip,\iSLAC}
\mbox{T.B. Moore\unskip,\iYALE}
\mbox{M. Morii\unskip,\iSLAC}
\mbox{D. Muller\unskip,\iSLAC}
\mbox{V. Murzin\unskip,\iMOSCOW}
\mbox{S. Narita\unskip,\iTOHO}
\mbox{U. Nauenberg\unskip,\iCOLO}
\mbox{H. Neal\unskip,\iYALE}
\mbox{G. Nesom\unskip,\iOXF}
\mbox{N. Oishi\unskip,\iNAGO}
\mbox{D. Onoprienko\unskip,\iTENN}
\mbox{L.S. Osborne\unskip,\iMIT}
\mbox{R.S. Panvini\unskip,\iVAND}
\mbox{C.H. Park\unskip,\iSOONG}
\mbox{I. Peruzzi\unskip,\iFRAS}
\mbox{M. Piccolo\unskip,\iFRAS}
\mbox{L. Piemontese\unskip,\iFERR}
\mbox{R.J. Plano\unskip,\iRUTG}
\mbox{R. Prepost\unskip,\iWISC}
\mbox{C.Y. Prescott\unskip,\iSLAC}
\mbox{B.N. Ratcliff\unskip,\iSLAC}
\mbox{J. Reidy\unskip,\iMISSI}
\mbox{P.L. Reinertsen\unskip,\iUCSC}
\mbox{L.S. Rochester\unskip,\iSLAC}
\mbox{P.C. Rowson\unskip,\iSLAC}
\mbox{J.J. Russell\unskip,\iSLAC}
\mbox{O.H. Saxton\unskip,\iSLAC}
\mbox{T. Schalk\unskip,\iUCSC}
\mbox{B.A. Schumm\unskip,\iUCSC}
\mbox{J. Schwiening\unskip,\iSLAC}
\mbox{V.V. Serbo\unskip,\iSLAC}
\mbox{G. Shapiro\unskip,\iLBL}
\mbox{N.B. Sinev\unskip,\iOREG}
\mbox{J.A. Snyder\unskip,\iYALE}
\mbox{H. Staengle\unskip,\iCSU}
\mbox{A. Stahl\unskip,\iSLAC}
\mbox{P. Stamer\unskip,\iRUTG}
\mbox{H. Steiner\unskip,\iLBL}
\mbox{D. Su\unskip,\iSLAC}
\mbox{F. Suekane\unskip,\iTOHO}
\mbox{A. Sugiyama\unskip,\iNAGO}
\mbox{A. Suzuki\unskip,\iNAGO}
\mbox{M. Swartz\unskip,\iJHU}
\mbox{F.E. Taylor\unskip,\iMIT}
\mbox{J. Thom\unskip,\iSLAC}
\mbox{E. Torrence\unskip,\iMIT}
\mbox{T. Usher\unskip,\iSLAC}
\mbox{J. Va'vra\unskip,\iSLAC}
\mbox{R. Verdier\unskip,\iMIT}
\mbox{D.L. Wagner\unskip,\iCOLO}
\mbox{A.P. Waite\unskip,\iSLAC}
\mbox{S. Walston\unskip,\iOREG}
\mbox{A.W. Weidemann\unskip,\iTENN}
\mbox{E.R. Weiss\unskip,\iWASH}
\mbox{J.S. Whitaker\unskip,\iBU}
\mbox{S.H. Williams\unskip,\iSLAC}
\mbox{S. Willocq\unskip,\iMASS}
\mbox{R.J. Wilson\unskip,\iCSU}
\mbox{W.J. Wisniewski\unskip,\iSLAC}
\mbox{J.L. Wittlin\unskip,\iMASS}
\mbox{M. Woods\unskip,\iSLAC}
\mbox{T.R. Wright\unskip,\iWISC}
\mbox{R.K. Yamamoto\unskip,\iMIT}
\mbox{J. Yashima\unskip,\iTOHO}
\mbox{S.J. Yellin\unskip,\iUCSB}
\mbox{C.C. Young\unskip,\iSLAC}
\mbox{H. Yuta\unskip.\iAOMORI}

\it
  \vskip \baselineskip                   % \bigskip did not work
  \centerline{(The SLD Collaboration)}   % include collaboration name
  \vskip \baselineskip
  \baselineskip=.75\baselineskip   % shrink the interline spacing
\iAOMORI
  Aomori University, Aomori, 030 Japan, \break
\iBRI
  University of Bristol, Bristol, United Kingdom, \break
\iBRUN
  Brunel University, Uxbridge, Middlesex, UB8 3PH United Kingdom, \break
\iBU
  Boston University, Boston, Massachusetts 02215, \break
\iCOLO
  University of Colorado, Boulder, Colorado 80309, \break
\iCSU
  Colorado State University, Ft. Collins, Colorado 80523, \break
\iFERR
  INFN Sezione di Ferrara and Universita di Ferrara, I-44100 Ferrara, Italy,
\break
\iFRAS
  INFN Laboratori Nazionali di Frascati, I-00044 Frascati, Italy, \break
\iJHU
  Johns Hopkins University,  Baltimore, Maryland 21218-2686, \break
\iLBL
  Lawrence Berkeley Laboratory, University of California, Berkeley, California
94720, \break
\iMASS
  University of Massachusetts, Amherst, Massachusetts 01003, \break
\iMISSI
  University of Mississippi, University, Mississippi 38677, \break
\iMIT
  Massachusetts Institute of Technology, Cambridge, Massachusetts 02139, \break
\iMOSCOW
  Institute of Nuclear Physics, Moscow State University, 119899 Moscow, Russia,
\break
\iNAGO
  Nagoya University, Chikusa-ku, Nagoya, 464 Japan, \break
\iOREG
  University of Oregon, Eugene, Oregon 97403, \break
\iOXF
  Oxford University, Oxford, OX1 3RH, United Kingdom, \break
\iPERU
  INFN Sezione di Perugia and Universita di Perugia, I-06100 Perugia, Italy,
\break
\iRAL
  Rutherford Appleton Laboratory, Chilton, Didcot, Oxon OX11 0QX United Kingdom,
\break
\iRUTG
  Rutgers University, Piscataway, New Jersey 08855, \break
\iSLAC
  Stanford Linear Accelerator Center, Stanford University, Stanford, California
94309, \break
\iSOONG
  Soongsil University, Seoul, Korea 156-743, \break
\iTENN
  University of Tennessee, Knoxville, Tennessee 37996, \break
\iTOHO
  Tohoku University, Sendai, 980 Japan, \break
\iUCSB
  University of California at Santa Barbara, Santa Barbara, California 93106,
\break
\iUCSC
  University of California at Santa Cruz, Santa Cruz, California 95064, \break
\iVAND
  Vanderbilt University, Nashville,Tennessee 37235, \break
\iWASH
  University of Washington, Seattle, Washington 98105, \break
\iWISC
  University of Wisconsin, Madison,Wisconsin 53706, \break
\iYALE
  Yale University, New Haven, Connecticut 06511. \break

\rm
%
%  }   % end of address list

\end{center}

\end{document}